\documentclass[aps,pre,twocolumn,showpacs]{revtex4}
\usepackage{graphicx}
\usepackage{amsmath}

\usepackage{units}
\usepackage{color}
\begin{document}
\title{Density and Correlation functions of vortex and saddle points in open billiard systems}
\author{R. H\"{o}hmann}
\author{U.~Kuhl}
\author{H.-J.~St\"{o}ckmann}
\affiliation{Fachbereich Physik der Philipps-Universit\"{a}t Marburg, Renthof 5, D-35032 Marburg, Germany}
\author{J. D.~Urbina$^{1,2}$}
\affiliation{$^1$ Department of Physics, Universidad Nacional de Colombia, Ciudad Universitaria, Bogota, Colombia}
\affiliation{$^2$Department of Physics, Regensburg Universit\"{a}t, 93047, Regensburg, Germany}
\author{M.R.~Dennis}
\affiliation{H H Wills Physics Laboratory, University of Bristol, Tyndall Avenue, Bristol BS8 1TL, UK}

\date{\today}

\begin{abstract}
We present microwave measurements for the density and spatial correlation of current critical points in an open billiard system, and compare them with the predictions of the Random Wave Model (RWM).
In particular, due to a novel improvement of the experimental set-up, we determine experimentally the spatial correlation of saddle points of the current field.
An asymptotic expression for the vortex-saddle and saddle-saddle correlation functions based on the RWM is derived, with experiment and theory agreeing well.
We also derive an expression for the density of saddle points in the presence of a straight boundary with general mixed boundary conditions in the RWM, and compare with experimental measurements of the vortex and saddle density in the vicinity of a straight wall satisfying Dirichlet conditions.
\end{abstract}

 \maketitle

\section{Introduction}

The use of Gaussian random functions to describe the spatial structure of complex physical systems has had a wide range of success, originating with
Rice's description of the random currents of shot noise \cite{ric45} and Longuet-Higgins' description of random water waves \cite{lon57a,lon57b}, more recently in such diverse fields as sound waves and acoustics \cite{ebe84b,tan07}, turbulence \cite{fri95}, optical speckle patterns \cite{goo07}, and the cosmic microwave background fluctuations \cite{lid00}.

In the realm of quantum wave physics, the same universality of the amplitude fluctuations has been conjectured in the spatial patterns of eigenfunctions in systems with classical (ray) chaotic dynamics \cite{ber77a}.
The analogy between the equations of a non-interacting two-dimensional electron gas and the electromagnetic modes of a microwave cavity \cite{stoe90,sri91,grae92a,so95,bar05a} (see Refs.~\onlinecite{stoe99,kuh07b} for reviews) allows a unified treatment in the language of quantum billiards, namely as solutions of the two-dimensional Helmholtz equation
\begin{equation}\label{eq:helm}
-(\partial_x^2+\partial_y^{2})\psi(\vec{r}) = k^{2}\psi(\vec{r})
\end{equation}
with wavenumber $k,$ energy $k^2,$ where $\vec{r} = (x,y).$
We therefore may study the properties of electron wavefunctions, which are difficult to access experimentally, by means of measurements using our microwave experimental setup.
Previously, insights from this analogy have had a strong impact on the theoretical study of coherent effects on electronic systems in the mesoscopic regime, where the spatial correlations of the electronic wavefunction are, besides the fluctuations of the energy spectra, the source of mesoscopic reproducible fluctuations (see Refs.~\onlinecite{mir00,tom1,gar} for recent examples).

Concerning fundamental questions, the measurements of complicated statistical measures (namely, averages over the experimentally constructed eigenfunctions of complicated functionals) are very stringent probes for the statistical assumptions upon which theoretical models of chaotic wavefunctions are based.
The primary such model is the so-called Random Wave Model (RWM), proposed by Berry \cite{ber77a}, based on the isotropic 2-dimensional random waves studied by Longuet-Higgins \cite{lon57b}.
The basic RWM is a statistically stationary isotropic solution of the two-dimensional Helmholtz equation (time-independent Schr\"{o}dinger equation), statistically invariant to translation and rotation.

The RWM predicts that the spatial fluctuations of eigenfuctions are Gaussian distributed, and this gives rise to characteristic morphological features (e.g.~Refs.~\onlinecite{ocon87,den07}).
So far, and largely due to the experimental possibility given by the microwave measurements, the assumption of Gaussian statistics has successfully passed very demanding tests.
To mention just two examples where very complicated functionals of the measured eigenfunctions are required and the RWM provide excellent results we have the distribution of current  \cite{seb99}, the intensity distribution in the transition from closed to open billiards \cite{kim05a} and the distribution of quantum stress tensor \cite{ber08b} (see Ref.~\onlinecite{kuh07b} for a recent review).

In this paper we address a different type of functional, based on the nodal properties of complex chaotic wavefunctions, and show again how the assumption of Gaussian statistics is strongly supported by the experimental results.

The features of the random eigenfunctions we study here are the critical points of the current density (hereafter current) associated with the complex wave $\psi.$
The current density is defined by
\begin{equation}\label{eq:j}
  \vec{j}(\vec{r}) \equiv \operatorname{Im}\psi(\vec{r})^* \nabla
  \psi(\vec{r})\,.
\end{equation}
In quantum-mechanical systems $\vec{j}(\vec{r})$ is representing the probability current density at position $\vec{r}$.
In quasi-two-dimensional electromagnetic microwave billiards there is a one-to-one correspondence of $\vec{j}(\vec{r})$ to the Poynting vector \cite{seb99}.

Since $\psi$ is assumed to satisfy the two-dimensional Helmholtz equation, the points where $\vec{j} = 0$ are of two types: vortices
of the flow (also known as circulations, wave dislocations, nodal points and phase singularities \cite{seb99,kuh07b,ber00b,sai01,den01b}), where $\psi = 0$ and about which the current swirls in a counterclockwise ($+$) or clockwise ($-$) sense, and saddle points (stagnation points), which are also saddle points of the phase $\arg \psi,$ and hyperbolic points in the current flow.
(The existence of phase extrema -- maxima or minima -- is prohibited by the Helmholtz equation \cite{den01b}.)
The topological Poincar\'{e} index of these types of points, describing the number of turns of $\vec{j}$ in a small circuit of the critical point, is $+1$ for vortices (regardless of the sense of circulation), and $-1$ for saddle points.
General arguments based on statistical isotropy demonstrate that there can be no net accumulation of topological charge, either in the sign of the vortices, or in the Poincar\'{e} index.
Therefore, the densities of positive and negative vortices must be equal, and the bulk vortex density must equal the bulk saddle density; calculations based on the RWM \cite{ber00b,sai01,den01b} give this vortex density as $k^2/4\pi.$
Knowledge of the positions of the critical points of the current vector field provides a skeleton on which the rest of the flow field is based.

An important deviation to the bulk RWM is caused by the presence of boundaries.
The interplay between spatial confinement and Gaussian fluctuations is by no means trivial, and it has even been claimed that in confined systems Gaussian fluctuations are valid only over very short distances \cite{gor02b}.
Substantial progress has been made recently in modifying the basic RWM to include boundary effects \cite{ber02b,ber02c,hor98a,whe05,bie02,urb06a,urb06b}, supported by numerical evidence \cite{ber05}, and there appears to be no reason to believe that the Gaussian assumption fails in the vicinity of a system's boundary.
Here, we present to our knowledge the first experimental measurements demonstrating the validity of boundary-adapted RWMs, based on Dirichlet conditions on an infinite straight wall.

\begin{figure}
\includegraphics[width=8.5cm]{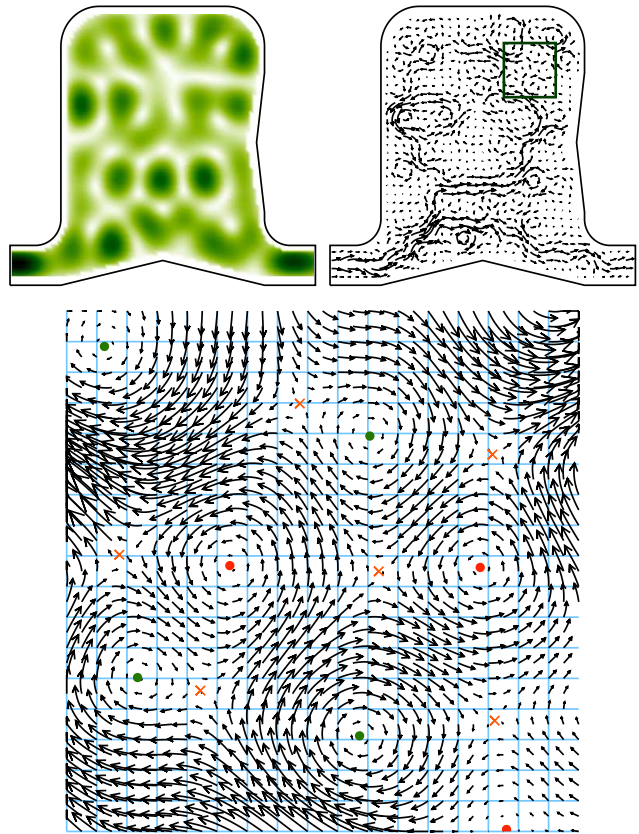}
\caption{\label{fig::data} (color online) Morphology of typical complex wavefunction $\psi$ in the open quantum billiard of our experiments.
(a) Modulus (intensity) $|\psi|^2;$ (b) Current flow $\operatorname{Im} \psi^* \nabla \psi$; (c) Blowup of (b), demonstrating the critical points we study.
The vortices are marked by dots, colored according to sense: counterclockwise (red), clockwise (green).
The saddle points are marked by crosses.
Our points of measurement are the crossing points in the background grid. }
\end{figure}

The intensity distributions in a two-dimensional chaotic microwave cavity, as shown in Fig.~\ref{fig::data}, are well-understood
\cite{kim05a}, although the distribution of current critical points has not previously been studied in detail.
In addition to the density fluctuations of vortices and saddle points against a Dirichlet boundary, we measure the vortex-vortex correlation function (including the case signed by circulation), and the vortex-saddle and saddle-saddle correlation functions, comparing against predictions of the RWM.

Although some of the theoretical predictions we compare with have been derived before (such as the vortex-vortex correlation functions
\cite{ber00b,sai01} and vortex density fluctuations against a straight wall \cite{ber02b,ber02c}), others are new.
In particular, we derive the density fluctuations of saddle points in the presence of an infinite straight boundary on which the wave satisfies mixed (Robin) boundary conditions, and, although we were unable to derive exact saddle correlation functions analytically, we have found large-$r$ asymptotic approximations to these functions.

The paper is organized as follows.
In section~\ref{ch::exp} we explain the experimental set-up and techniques used to locate the critical points in the microwave cavity.
General definitions and properties of critical points of the current associated with the Helmholtz equation, as well as the expressions for densities and correlations, are described in section~\ref{ch::crit}.
This is followed by section~\ref{ch::Averages}, which describes our RWM calculations: the model is introduced in subsection~\ref{sub:RWM}, and details follow for bulk correlation functions (subsection \ref{sub:bulk}) and densities near a straight boundary satisfying Robin conditions (subsection~\ref{sub:boundaries}).
The experimental results are compared with the theoretical predictions in section~\ref{ch::comparison}.

\section{Experimental setup}
\label{ch::exp}

We here report on the measurement of correlations functions of saddle and vortex points in an open billiard system including effects of the boundaries.
The basic principles of the experiment can be found in  Ref.~\onlinecite{kuh00b}.
We used a rounded rectangular cavity\,(21\,cm$\times$\,16\,cm) coupled to two wave guides of width 3\,cm with a cut-off frequency at $\nu_T=5$\,GHz.
To break the symmetry and to block direct transport, two triangular obstacles with a length of 12\,cm and a height of 1\,cm were placed in the resonator.
Absorbers were placed at the end of the leads to avoid reflection.
We scanned the billiard on a square grid of 2.5\,mm with a movable antenna $A_1$ and measured transmission $S_{12}$ in the range of 4\,-\,18\,GHz from a fixed antenna $A_2$ in the end of the right lead.
The fixed antenna had a metallic core of diameter 1\,mm and a Teflon coating while the probe antenna $A_1$ was a thin wire of diameter 0.2\,mm to minimize the leakage current.
The lengths of the antenna $A_1$ and antenna $A_2$ were 4 and 5\,mm respectively.

For microwave frequencies $\nu < c/2d=18.75\,$GHz, where $c$ is the velocity of light and $d$ is the resonator height, the billiard is quasi-two-dimensional.
In this regime there is an exact correspondence between electrodynamics and quantum mechanics, where the component of the electric field perpendicular to the plane of the microwave billiard $E_z$ corresponds to the quantum-mechanical wave function $\psi$.
Figure~\ref{fig::data} shows typical intensity and current patterns thus obtained.
Additionally a zoom of the flow pattern is shown to visualize the structure and showing the different types of critical points in the flow.

We previously have reported results on vortex pair correlation functions and nearest neighbor distance distributions [\onlinecite{kim03b,bar02}] for vortex points.
Through improvements in data analysis, we have been able to study the saddle points of the current in addition to the vortices, and significantly reduce the effects of noise in our measured correlation functions.
These improvements are sketched in the following.
We have increased the spatial resolution by a factor of two compared to our previous measurements before and we now use additionally a bilinear interpolation method for the individual components of the flow.
Using the bilinear interpolation we estimate the nodal lines of the individual flow components, enabling us to get the exact position for vortex and saddle points within this approximation.
The extraction of critical point locations is now fully automatic, allowing large samples of data to be analyzed.
One can see the effect of these improvements in the better results for the pair correlation functions (good agreements also for small $kr$) and the charge correlation function which we can present here for the first time.

\section{Current critical point densities}
\label{ch::crit}

In this section and following, we assume that the complex wavefunction $\psi(\vec{r}) = \xi(\vec{r})+{\rm i} \eta(\vec{r})$ has no particular symmetries or properties, beyond satisfying Eq.~(\ref{eq:helm}).
The current $\vec{j},$ from Eq.~(\ref{eq:j}), can therefore be written
\begin{equation}\label{eq:j1}
 \vec{j} = ( \xi \eta_x - \eta \xi_x, \xi \eta_y - \eta \xi_y ).
\end{equation}
$\vec{j} = 0$ at vortices, where $\psi = \xi = \eta = 0,$ and at saddle points, where $\xi/\eta = \xi_x/\eta_x = \xi_y/\eta_y$ (if $\eta = 0,$ there is equality between the reciprocals of these terms).

The quantity which distinguishes vortices from saddles is the Jacobian
\begin{equation}\label{eq:jjac}
 \mathcal{J} =  \partial_x j_x \partial_y j_y - \partial_y j_x \partial_x j_y,
\end{equation}
which is positive at vortices, and negative at saddles.
Since
$\psi$ satisfies Eq.~(\ref{eq:helm}), $\mathcal{J}$ separates into two contributions \cite{den01b},
\begin{equation}\label{eq:jsep}
 \mathcal{J} = \mathcal{J}_{\mathrm{v}} - \mathcal{J}_{\mathrm{s}},
\end{equation}
where
\begin{eqnarray}
 \mathcal{J}_{\mathrm{v}} & \equiv & (\xi_x \eta_y - \xi_y \eta_x)^2, \label{eq:jvort} \\
 \mathcal{J}_{\mathrm{s}} & \equiv & \frac{1}{2} (\xi \eta_{xx} - \eta \xi_{xx})^2 + \frac{1}{2}(\xi \eta_{yy} - \eta \xi_{yy})^2  \nonumber \\
 & & + (\xi \eta_{xy} - \eta \xi_{xy})^2. \label{eq:jsad}
\end{eqnarray}
Obviously, $\mathcal{J}_{\mathrm{v}}=0$ at saddle points, and $\mathcal{J}_{\mathrm{s}} = 0$ at vortices.
This fact, combined with positive-definiteness of the two parts of the Jacobian, implies that $|\mathcal{J} | = \mathcal{J}_{\mathrm{v}}$ at vortices, and $|\mathcal{J} | = \mathcal{J}_{\mathrm{s}}$ at saddle points.

These quantities can be used to define functions which find critical points (vortices or saddles) at position $\vec{r}$.
The density of critical points, with a unit $\delta$-function at each zero point of
$\vec{j}$, is given by
\begin{equation}
  D_{\mathrm{crit}}(\vec{r}) \equiv \sum_{ \{\vec{r} \, : \, \vec{j}(\vec{r}) = 0\} } \delta^2(\vec{r}) = \delta^2(\vec{j}(\vec{r})) |\mathcal{J}^\prime(\vec{r})|.
  \label{eq:Dcrit}
\end{equation}
By the separation of $\mathcal{J}$ above, this gives the saddle density \cite{den01b},
\begin{equation}
  D_{\mathrm{s}}(\vec{r}) \equiv \delta^2(\vec{j}(\vec{r})) \mathcal{J}_{\mathrm{s}}(\vec{r}).
  \label{eq:Dsad}
\end{equation}
and the vortex density \cite{ber00b,sai01,den01b},
\begin{equation}
  D_{\mathrm{v}}(\vec{r}) \equiv \delta^2(\vec{j}(\vec{r})) \mathcal{J}_{\mathrm{v}}(\vec{r}) = \delta(\xi)\delta(\eta) |\xi_x \eta_y - \xi_y \eta_x|.
  \label{eq:Dv}
\end{equation}
The vortex sign (sense of circulation) is given by
\begin{equation}
   \mathcal{S} \equiv \operatorname{sign}(\xi_x \eta_y - \xi_y \eta_x),
   \label{eq:vortsign}
\end{equation}
so removing modulus signs gives the signed vortex density.

The number and location of critical points for a given field must be founded by explicitly solving the set of equations $\vec{j}(\vec{r}) = 0$.
This of course requires the precise knowledge of the spatial dependence of the particular solution $\psi(\vec{r})$ in which we are interested.
The task of solving the Helmholtz equation in cases where the geometry of the confinement (transversal section of the waveguide) is such that Eq.~(\ref{eq:helm}) is not separable is usually very demanding.
This makes the function-by-function study of current morphology almost impossible.

A suitable way to overcome this complication is to use a statistical
approach.
This idea is based on the strong uniformity of the solutions of the Helmholtz equation with non-integrable geometries (see for example Fig.~\ref{fig::data}), indicating that their main properties actually depend on far fewer parameters than the full solution itself.
We therefore consider, instead of a given set of solutions of the Helmholtz equation, an {\it ensemble} of fields.
This ensemble will be constructed in such a way that the most general and basic properties of the exact solutions are respected, in the hope that these general properties suffice to fix the morphology fluctuations.
The ensemble we choose is the usual random wave model (RWM) discussed in the Introduction, or the boundary-adapted model of Refs.~\onlinecite{ber02b,ber02c}.

Delaying on the appropriate definition of the RWM until the next section, we merely write the average over the ensemble $\langle \cdots \rangle$.
In this paper, we compare theoretical RWM predictions and experimental measurements of the average density fluctuations
\begin{equation}
  \rho_{\mathrm{\alpha}}(\vec{r}) = \frac{4\pi}{k^2}\langle D_{\alpha}(\vec{r}) \rangle,
  \label{eq:densities}
\end{equation}
and 2-point correlations
\begin{equation}
  g_{\alpha\beta}(\vec{r}_2,\vec{r}_1) = \left(\frac{4\pi}{k^2}\right)^2 \langle D_{\alpha}(\vec{r}_2) D_{\beta}(\vec{r}_1) \rangle,
  \label{eq:correlations}
\end{equation}
where $\alpha,\beta$ are v,s.
These expressions have been normalized against the bulk average vortex density (and saddle density) $k^2/4\pi$ \cite{ber00b,sai01,den01b}.
Statistical symmetries in the RWMs will mean that the densities $\rho$ and correlations $g$ have simpler functional dependence.
2-point correlation functions can also be considered which take topological signs into account, such as the vortex topological charge correlation function
\begin{equation}\label{eq:gQdef}
  g_Q(\vec{r}_1,\vec{r}_2) = \frac{\langle D_{\mathrm{v}}(\vec{r}_1) \mathcal{S}(\vec{r}_1)) D_{\mathrm{v}}(\vec{r}_2) \mathcal{S}(\vec{r}_2)) \rangle}{(k^2/4\pi)^2},
\end{equation}
where the effect of the signum $\mathcal{S}$ functions of Eq.~(\ref{eq:vortsign}) is to negate the modulus signs in $D_{\mathrm{v}}$; the vortices are signed by their sense of circulation.
Current critical point correlation functions can be written down in terms of the correlation functions $g_{\alpha\beta}.$
If the 2-point critical point correlation function is denoted $g_{\mathrm{crit}},$ and $g_I$ the function signed by Poincar\'{e} index (positive for vortices, negative for saddles), we have
\begin{eqnarray}
  g_{\mathrm{crit}}(\vec{r}_1,\vec{r}_2) & = & \frac{1}{4}\left(g_{\mathrm{vv}}(\vec{r}_1,\vec{r}_2) + g_{\mathrm{ss}}(\vec{r}_1,\vec{r}_2) + 2 g_{\mathrm{vs}}(\vec{r}_1,\vec{r}_2) \right), \nonumber \\
  \label{eq:gcritdef} \\
  g_{I}(\vec{r}_1,\vec{r}_2) & = & \frac{1}{4}\left(g_{\mathrm{vv}}(\vec{r}_1,\vec{r}_2) + g_{\mathrm{ss}}(\vec{r}_1,\vec{r}_2) - 2 g_{\mathrm{vs}}(\vec{r}_1,\vec{r}_2) \right). \nonumber \\
  \label{eq:gIdef}
\end{eqnarray}

\section{Average densities and correlations of critical points within the random wave model}
\label{ch::Averages}

\subsection{RWM: basic definition and field correlations}
\label{sub:RWM}

The random wave model assumes that the wave field, satisfying the Helmholtz equation (\ref{eq:helm}), is a superposition of infinitely many complex plane waves with wavenumber $k$ with uniformly random directions and phases; the real and imaginary parts of the field are therefore assumed to be statistically independent.
These assumptions, while appropriate to our experimental open billiards, which lack time reversal symmetry, do not apply in the transition from open to closed systems \cite{kuh07b}.
The RWM is ergodic in the bulk -- spatial averages are equivalent to ensemble averages.
We will describe the boundary-adapted random wave model at the end of this subsection.

The central limit theorem ensures that, in the limit of infinitely many superposed plane waves, the probability density function of the value of the wave at each point has a complex circular Gaussian distribution \cite{goo07,ber00b}.
Furthermore, the distribution of all derivatives of the field are also Gaussian random variables, which may have nonvanishing correlations with each other and the original field.

The assumption that the field and its derivatives possess multivariate Gaussian statistics implies that, for a functional $\mathcal{F}[\vec{u}],$ depending on the field and its derivatives at possibly different points, we have
\begin{equation}\label{eq:gauss}
  \langle \mathcal{F} \rangle = \frac{1}{\sqrt{(2 \pi)^{n} \det \bf{M}}}\int_{-\infty}^{\infty} \mathcal{F}[\vec{u}] {\rm e}^{-\frac{1}{2} \vec{u}\cdot {\bf M}^{-1}\cdot \vec{u}} \mathrm{d}^n\vec{u}.
\end{equation}
where $\vec{u}$ is an $n$-dimensional vector consisting of the relevant Gaussian random fields $\xi(\vec{r}_1),\partial_y \eta(\vec{r}_2),$ etc.~appearing in $\mathcal{F},$ and $\bf{M}$ is the $n \times n$ matrix of correlations with entries $M_{i,j},$
\begin{equation}\label{eq:mij}
  M_{i,j}=\langle u_{i} u_{j} \rangle.
\end{equation}

Calculating densities of morphological features in the RWM is therefore reduced to a Gaussian integral, whose difficulty depends on the complexity of the functional $\mathcal{F}.$
For instance, the average density of vortices or saddles in the bulk isotropic random waves can be calculated with $\mathcal{F} = D_{\mathrm{v}}$ or $D_{\mathrm{s}}$ from Eqs.~(\ref{eq:Dsad}), (\ref{eq:Dv}); as discussed previously these are known to be equal constants, with value $k^2/4\pi$ \cite{den01b}.
In this paper, we concentrate on two specific types of functionals $\mathcal{F}.$

In subsection~\ref{sub:bulk}, we consider two-point correlations $g_{\alpha\beta}$ in the bulk isotropic random wave model, where $\mathcal{F}$ is given by $D_{\mathrm{\alpha}}(\vec{r}_1) D_{\mathrm{\beta}}(\vec{r}_2),$ with $\alpha, \beta = \mathrm{v}, \mathrm{s}.$
These expressions are the average densities of vortices or saddles at two points, depending only on the scaled distance
\begin{equation}\label{eq:Rdef}
  R \equiv k|\vec{r}_2 - \vec{r}_1|,
\end{equation}
by isotropy.
These critical point correlation functions depend only on the 2-point field correlation function, given by
\begin{equation}\label{eq:Rralpha}
  C(R)=\frac{1}{2} \langle \psi(\vec{r}_1) \psi^{*}(\vec{r}_2) \rangle = \langle \xi(\vec{r}_1) \xi(\vec{r}_2) \rangle = \langle \eta(\vec{r}_1) \eta(\vec{r}_2) \rangle.
\end{equation}
All 2-point correlation functions of derivatives of the field can be expressed in terms of derivatives of $C(R).$
It is well known \cite{ber77a} that the field correlation function of the 2-dimensional isotropic RWM is given by the Bessel function
\begin{equation}\label{eq:Cdef}
  C(R) = J_{0}(R).
\end{equation}
Much of our argument will be based on asymptotic approximations for large $R,$ in which, to leading order,
\begin{equation}\label{eq:Cas}
  C(R) \stackrel{R \gg 1}{\sim} \sqrt{\frac{2}{\pi R}} \cos(R - \pi/4).
\end{equation}

\begin{figure}
\includegraphics[width=8cm]{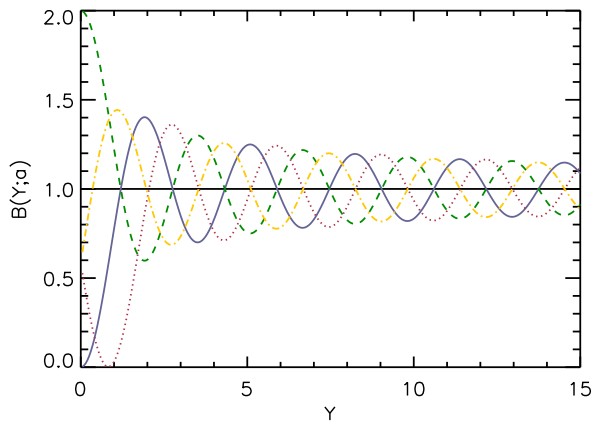}
\caption{\label{fig:wallfield} (color online)
 The normalized mean field intensity $B(Y;a),$ plotted against $Y$ for various choices of Robin parameter $a$: $a = 0,$ i.\,e.~Dirichlet conditions (solid, blue); $a = \pi/4$ (dotted, purple); $a = \pi/2,$ i.\,e.~Neumann conditions (dashed-dotted, yellow); $a = -\pi/4$ (dashed, green).
 The black line is at 1 (the asymptotic limit for $Y \to \infty$)}.

\end{figure}

The second type of functional we consider is based on the vortex and saddle densities $D_{\mathrm{v}}, D_{\mathrm{s}}$ in the so-called boundary adapted random wave model.
In this model, the wave with $y\ge 0$ is assumed to satisfy a homogeneous boundary condition along the infinite straight line $y = 0.$
As with the correlation function above, we will use dimensionless Cartesian coordinates
\begin{equation}\label{eq:XYdef}
  X \equiv k x, \quad Y \equiv k y.
\end{equation}
Although our experimental data are for Dirichlet conditions only ($\psi(X,0) = 0$), our discussion will be framed in terms of the most general boundary conditions, namely mixed (Robin) conditions
\begin{equation}\label{eq:robin}
  \psi(X,0) \cos a  + \partial_Y \psi(X,0) \sin a  = 0,
\end{equation}
where $a$ is a constant with $0 \le a < \pi.$
Dirichlet conditions correspond to $a = 0,$ and Neumann to $a = \pi/2.$
Berry and Ishio \cite{ber02c} constructed a natural RWM satisfying Eq.~(\ref{eq:robin}), and  considered the vortex density as a function of distance $Y$ from the boundary, and general $a,$ generalizing previous work \cite{ber02b} for Dirichlet and Neumann conditions.
We will calculate the corresponding $Y$-dependent saddle point density in subsection~\ref{sub:boundaries}.
The critical point density calculations depend on the $Y$-dependent, 1-point quadratic field correlation
\begin{equation}\label{eq:Bdef}
  B(Y;a) \equiv  \langle \xi(X,Y)^2 \rangle = \langle \eta(X,Y)^2 \rangle
\end{equation}
For general $a,$ the function $B(Y;a)$ for the mixed boundary condition RWM of Ref.~\onlinecite{ber02c} cannot be expressed in terms of elementary functions, although it has a straightforward integral representation:
\begin{eqnarray}\label{eq:Bint}
  & &  B(Y;a) = 1 - \frac{\pi}{2} \int_0^{\pi/2} \mathrm{d}\theta  \\
  & & \quad \times \frac{(1 - \tau \sin^2\theta) \cos(2Y \sin\theta) + 2 \tau \sin \theta \sin(2Y \sin \theta)} {1+\tau \sin^2\theta} \nonumber
\end{eqnarray}
with $\tau = \tan a.$
For Dirichlet and Neumann conditions, $B$ can be expressed in terms of the Bessel function $J_0,$
\begin{equation}\label{eq:BDN}
  B\left(Y;\begin{array}{c} 0\\ \pi/2 \end{array}\right) = 1 \mp J_0(2Y).
\end{equation}
It is straightforward to find an asymptotic approximation for $B(Y;a),$
\begin{equation}\label{eq:Basym}
  B(Y;a) \stackrel{{Y\gg 1}}{\sim} 1 - \frac{1}{\sqrt{\pi Y}} \cos(2Y - 2a -\pi/4),
\end{equation}
consistent with Eq.~(\ref{eq:BDN}).
The field intensity fluctuation $B(Y;a)$ is plotted as a function of $Y$ for various choices of $a$
in Fig.~\ref{fig:wallfield}.
We mention that the result in Eq.~(\ref{eq:Basym}) is also found by using the semiclassical approximation for the two-point correlation function in the presence of boundaries with Robin boundary conditions.
In this case, the parameter $a$ enters through the semiclassical phase, as explained in \cite{urb06b}.

\subsection{Spatial correlations of current vortices and saddles: bulk results}
\label{sub:bulk}

Correlations of vortices and related objects have been the subject of much study in the isotropic RWM.
In particular, the signed vortex-vortex correlation function $g_Q(R)$ defined in Eq.~(\ref{eq:gQdef}), is known to have a remarkably simple form \cite{ber00b,hal81,fre98,fol03b,fol03a,den03,wil04}
\begin{equation}
  g_{Q}(R) = \frac{4}{R}\frac{\mathrm{d}}{\mathrm{d}R}\left[\frac{\mathrm{d}\arcsin(J_0(R))}{\mathrm{d}R}\right]^2.
  \label{eq:gQ}
\end{equation}
This equation (with $J_0$ replaced by a suitable 2-point function) holds for general isotropic Gaussian random fields, not just the Bessel-correlated random wave model.
On account of the isotropy of the distribution in the phase of the field $\psi,$ $g_Q$ satisfies the `topological charge screening relation' (ignoring the self-interaction at $R = 0$) \cite{ber00b,hal81,fre98,fol03b,fol03a,den03,wil04}
\begin{equation}
  \frac{1}{2} \int_0^{\infty} \mathrm{d}R \, R g_{Q}(R) = -1.
  \label{eq:screening}
\end{equation}
For $R \gg 1,$ $g_{Q}(R) \sim 8 \cos(2R)/\pi R^2.$
The oscillation period of $g_Q(R)$ is twice that of the correlation function $C(R)$ since there are two nodes per oscillation of $J_0.$

However, the unsigned correlations we emphasize here, including saddle correlations, do not have such a simple form.
The vortex-vortex correlation function $g_{\mathbf{vv}}(R)$ was computed exactly in Refs.~\onlinecite{ber00b,sai01}, and is written down in Ref.~\onlinecite{den01c} Eqs.~(32)-(35) (also see Ref.~\onlinecite{den01a} p.~83) as a complicated function involving various elliptic integrals; using similar methods involving computer algebra, our attempts to extract $g_{\mathrm{vs}}(R)$ and $g_{\mathrm{ss}}(R)$ analytically were unsuccessful.
We therefore developed a strategy, based on an asymptotic expansion of the correlation matrix through the asymptotic expansion of the Bessel function, as indicated in the subsection~\ref{sub:RWM}, to extract asymptotic approximations for these correlation functions (including $g_{\mathrm{vv}}(R)$) to compare with the experimental data.
Details are given in appendix \ref{app:asymptotic}.
Our asymptotic argument is similar to the asymptotic 2-point correlation function for gradient saddles in computed in the real RWM \cite{fol}, based on functional differentiation.

In order to compute the saddle-saddle spatial correlation we need to consider as degrees of freedom the field and its derivatives at two different points, $\vec{r}_1, \vec{r}_2$.
The basic idea of the asymptotic method is that in this case the correlation matrix in Eq.~(\ref{eq:gauss}) can be unambiguously separated into two contributions.
On one side, we have the correlations of the degrees of freedom at the same point, which is of course independent of $R = k |\vec{r}_1 - \vec{r}_2|$.
All the spatial dependence of the correlation is encoded in the elements of the correlation matrix relating the fields at different points.

\begin{figure}
\includegraphics[width=8cm]{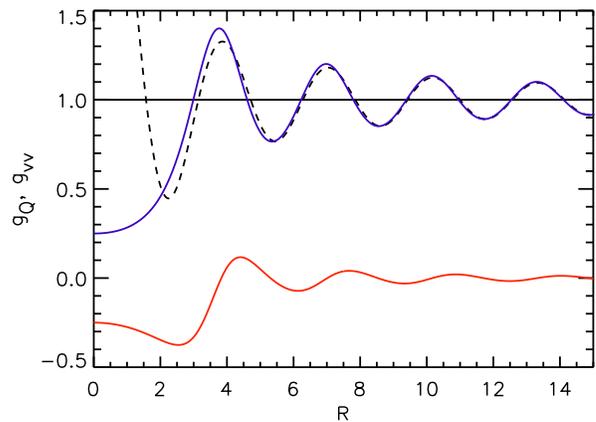}
\caption{\label{fig:gvvgq} (color online)
 Analytic 2-point correlation functions $g_{\mathrm{vv}}(R)$ (red) and $g_Q(R)$ (green), plotted against $R.$
 The black line is the asymptotic value of 1 for $g_{\mathrm{vv}}(R)$ as $R \to \infty$.
 The analytic functions compare well with their asymptotic approximations (dashed curves) for $R \gtrsim 4.$
}
\end{figure}

Once this separation of the spatial dependence of the correlation matrix is realized, the asymptotic expansion of the two-point correlation function Eq.~(\ref{eq:Cas}) naturally leads to an expansion of any arbitrary average in powers of the small parameter $1/\sqrt{R}$.
Obviously, this expansion will fail for short distances (comparison with the experimental results show that in practice the approximation is good already for $R \gtrsim 3$).
Omitting further details of the derivation, the relevant asymptotic approximations to order $O(R^{-1})$ of the RWM vortex-vortex, vortex-saddle, and saddle-saddle pair correlations are
\begin{eqnarray}
  g_{\mathrm{vv}}(R) & \sim & 1 + \frac{4 \sin 2R}{\pi R},
  \label{eq:asymgvv} \\
  g_{\mathrm{vs}}(R) & \sim & 1 - \frac{4 \sin 2R}{\pi R},
  \label{eq:asymgvs} \\
  g_{\mathrm{ss}}(R) & \sim & 1 + \frac{4 \sin 2R}{\pi R}.
  \label{eq:asymgss}
\end{eqnarray}
Fig.~\ref{fig:gvvgq} is a plot of $g_{\mathrm{vv}}(R),$ computed analytically \cite{den01c}, and from the asymptotic form (\ref{eq:asymgvv}).
The 2-point correlation functions for vortices of the same sign $g_{++}(R) = g_{--}(R) \equiv \frac{1}{2}(g_{\mathrm{vv}}(R) + g_Q(R))$ and opposite sign $g_{+-}(R) \equiv \frac{1}{2}(g_{\mathrm{vv}}(R) - g_Q(R))$ oscillate in phase \cite{den07} since $g_Q \sim O(R^{-2})$ decays more swiftly than $g_{\mathrm{vv}}(R).$
These equations demonstrate that critical points, that is vortices (whose Poincar\'{e} index is $+1$) and saddles (with index $-1$) oscillate out of phase.

Eqs.~(\ref{eq:asymgvv})-(\ref{eq:asymgss}) can be used to estimate asymptotically the critical point 2-point functions $g_{\mathrm{crit}}(R),$ and its Poincar\'{e} index-signed analogue $g_{I}(R)$:
\begin{eqnarray}
  g_{\mathrm{crit}}(R) & \sim & 1 + O(R^{-2}),
  \label{eq:asymgcrit} \\
  g_I(R) & \sim & 4 \sin(2R)/\pi R. \label{eq:asymgI}
\end{eqnarray}
$g_I(R)$ decays rather slowly, in contrast to the long range correlations of topological charges of other RWMs, such as critical points of the gradient in the real RWM \cite{fol03a,den03}, which decay to leading order like $O(R^{-3}).$
It would therefore be interesting to establish whether Poincar\'{e} index satisfies a screening relation analogous to Eq.~(\ref{eq:screening}), since convergence in the integral is marginal.

\subsection{Densities of critical points near mixed boundaries}
\label{sub:boundaries}

In the boundary-adapted RWM, the density of vortices and saddles oscillates with distance $Y$ from the boundary, just as the square
field does (e.g.~Eq.~(\ref{eq:Basym})).
However, with the Robin RWM of Ref.~\onlinecite{ber02c}, the form of the vortex density function depends only on the function $B(Y;a)$ of Eq.~(\ref{eq:Bdef}); the result of the actual Gaussian integral is independent of the value of parameter $a.$
The entries of the correlation matrix, of course, do depend on $a,$ and all may be written as linear combinations of $B(Y;a)$ and its derivatives with respect to $Y.$

Therefore, the mean vortex density in the boundary-adapted RWM, is
\cite{ber02b}
\begin{equation}\label{eq:rhovba}
  \rho_{\mathrm{v}}(Y;a) = \frac{\sqrt{4B+B''-1} \sqrt{B(2+B'')-B'^2}}{2 B^{3/2}},
\end{equation}
where dependence of $B$ on $Y$ and $a$ is suppressed and the prime denotes the partial derivative with respect to $Y$.
The density is normalized with respect to the bulk density, so
$\lim_{Y\to\infty}\rho_{\mathrm{v}}(Y;a) = 1.$
This equation is the same as Ref.~\onlinecite{ber02b}, Eq.~(40), with correlation matrix elements replaced by appropriate functions of $B.$

Since the saddle density $D_{\mathrm{s}}(\vec{r})$ of Eq.~(\ref{eq:Dsad}) does not involve modulus signs, the calculation of the average saddle density in the boundary-adapted RWM uses straightforward Gaussian integration techniques, as outlined in Appendix \ref{app:dervrhosba}.
The resulting density is
\begin{eqnarray}
  \rho_{\mathrm{s}}(Y;a) & = & \frac{B^{1/2}}{(4B+B''-2)^{3/2}(B(2+B'')-B'^2)^{3/2}} \nonumber \\
  & & \quad \times \left( 16 - 64 B + 64 B^2 + 16 B'^2 - 64 B B'^2  \right. \nonumber \\
  & & \qquad \; + 16 B'^4 - 16 B'' + 64 B^2 B'' - 32 B B'^2 B'' \nonumber \\
  & & \qquad \; + 16 B B''^2 + 16 B^2 B''^2 - 4 B'^2 B''^2  \nonumber \\
  & & \qquad \; + 4 B''^3 - B''^4 + 8 B' B''' - 32 B B' B'''  \nonumber \\
  & & \qquad \; + 8 B'^3 B''' - 8 B' B'' B''' + 2 B' B''^2 B'''  \nonumber \\
  & & \qquad \; - 4 B^2 B'''^2 + B'^2 B'''^2 - 2 B B'' B'''^2   \nonumber \\
  & & \qquad \; - 8 B B'''' + 16 B^2 B''''+ 4 B'^2 B''''   \nonumber \\
  & & \qquad \; - 8 B B'^2 B'''' + 8 B^2 B'' B'''' \nonumber \\
  & & \qquad \; \left. - 2 B'^2 B'' B'''' + 2 B B''^2 B'''' \right)
  \label{eq:rhosba}
\end{eqnarray}

\begin{figure}
\includegraphics[width=7cm]{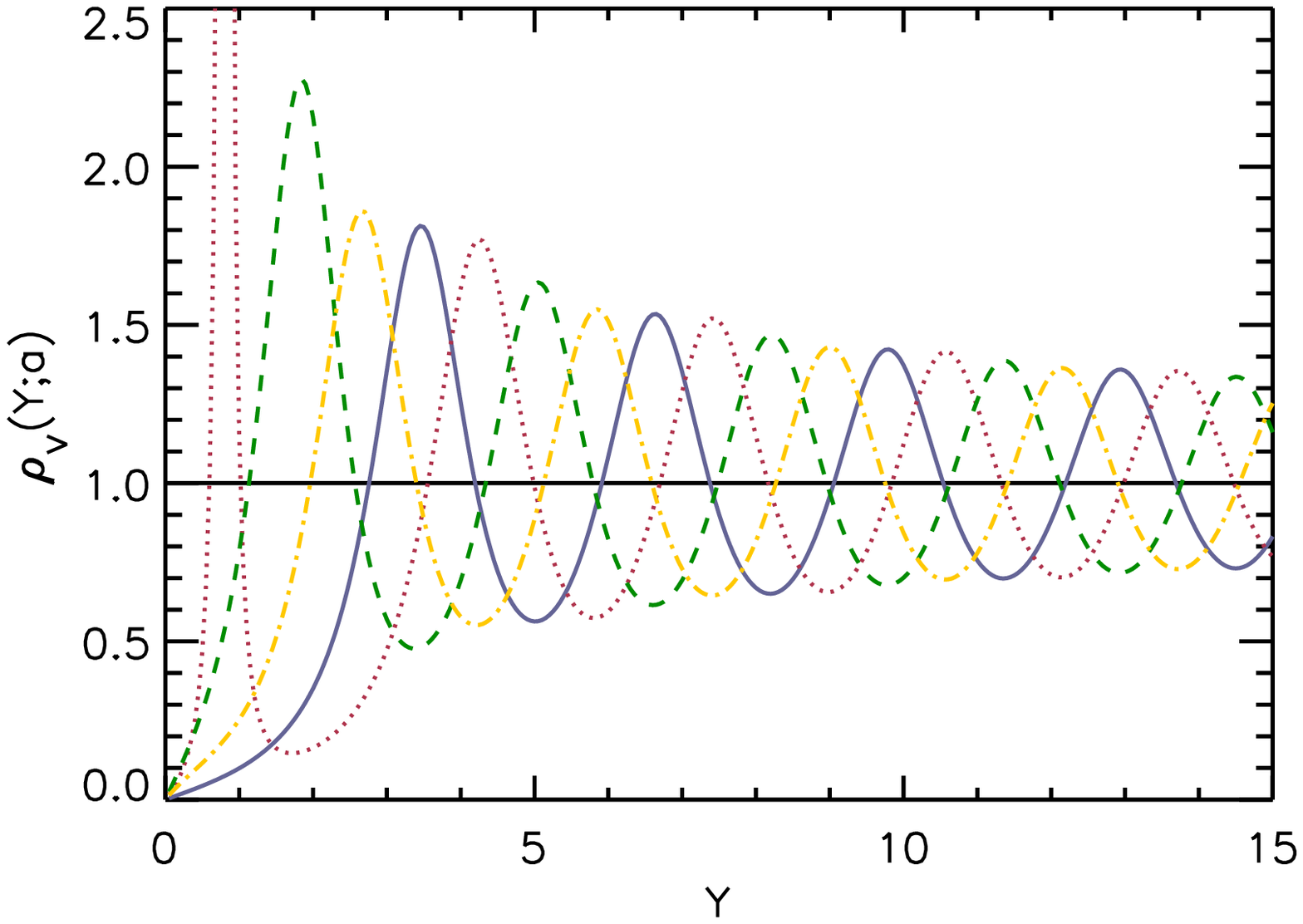}
\includegraphics[width=7cm]{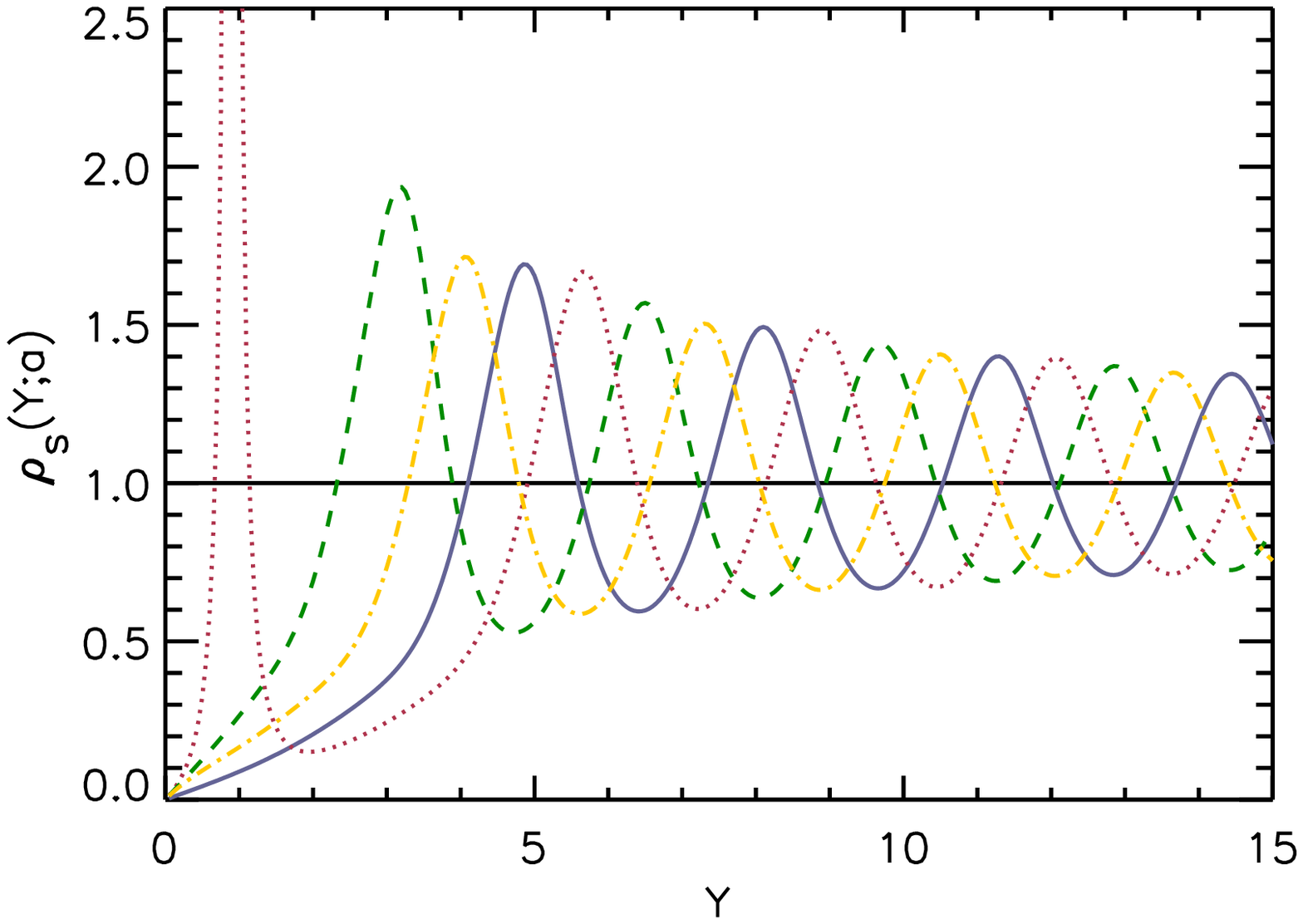}
\includegraphics[width=7cm]{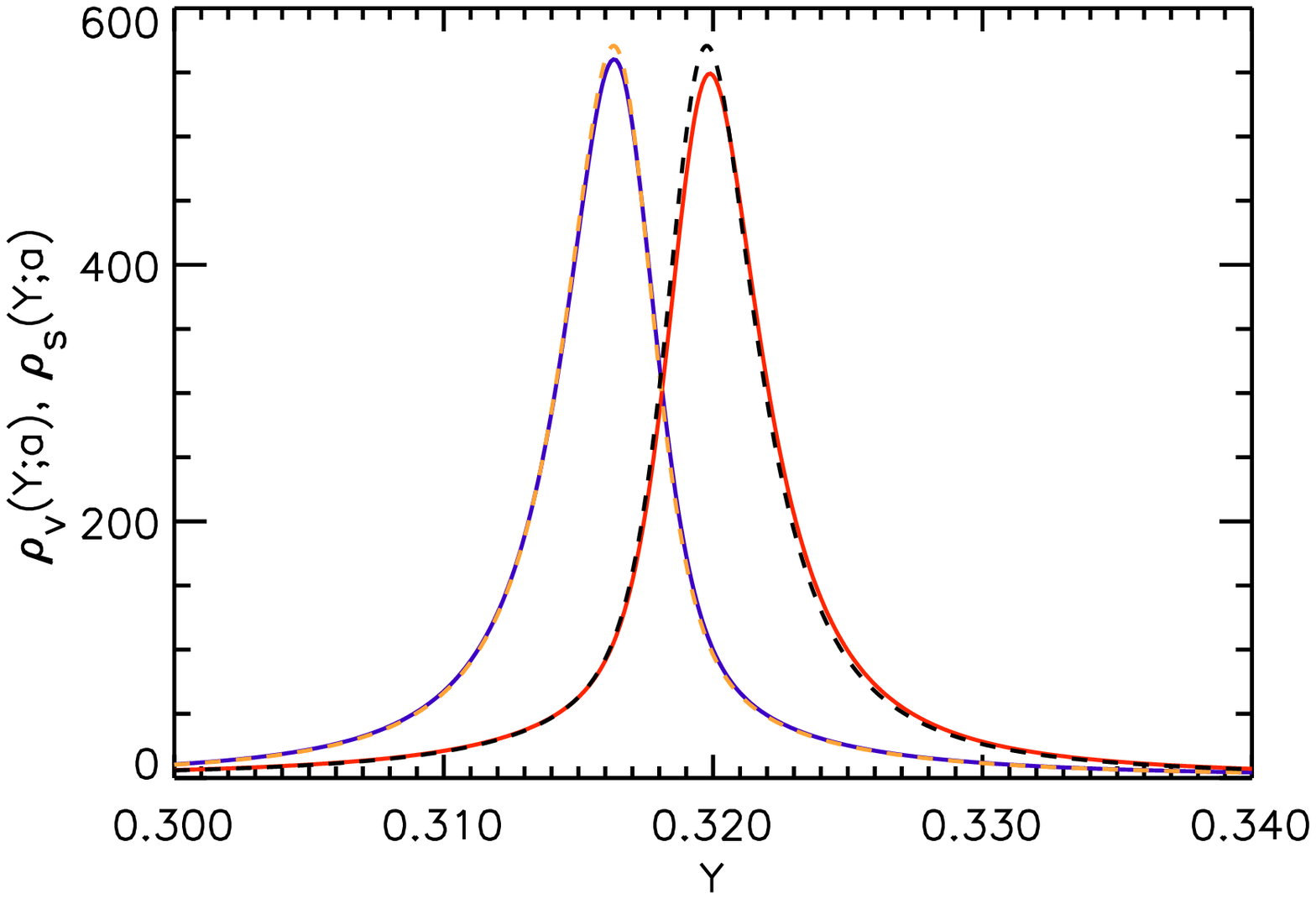}
\caption{\label{fig:wall} (color online)
 Density oscillations of critical points as a function of distance $Y$ from a wall satisfying mixed Robin conditions: (a) $\rho_{\mathrm{v}}(Y;a);$ (b) $\rho_{\mathrm{s}}(Y;a).$
 The colors represent the same choices of $a$ as in Fig.~\ref{fig:wallfield}, the black line is at 1 (the asymptotic limit for $Y \to \infty$); the two densities are clearly out of phase for $Y \gtrsim 4.$
 (c) Peak in the densities for $a = \pi/10:$ $\rho_{\mathrm{v}}$ (red curve), $\rho_{\mathrm{s}}$ (green curve), with the dashed lines the small-$a$ forms of Eqs.~(\ref{eq:rhovsmalla}), (\ref{eq:rhossmalla}).
}
\end{figure}

Asymptotically, for $Y \gg 1$ we find
\begin{eqnarray}
  \rho_{\mathrm{v}}(Y;a) & \sim & 1 + \frac{2\cos(2(Y - a) -\frac{\pi}{4})}{\sqrt{\pi Y}} + \frac{1+5\sin(4(Y-a))}{4\pi Y}, \nonumber \\
  \label{eq:rhovasym} \\
  \rho_{\mathrm{s}}(Y;a) & \sim & 1 - \frac{2\cos(2(Y - a) -\frac{\pi}{4})}{\sqrt{\pi Y}} + \frac{1+5\sin(4(Y-a))}{4\pi Y} \nonumber \\
  \label{eq:rhosasym}
\end{eqnarray}
(Eq.~(\ref{eq:rhovasym}) was demonstrated in Ref.~\onlinecite{ber02c}).
Therefore the leading order oscillations in vortex and saddle densities are exactly out of phase, as in the 2-point correlation functions discussed in the previous subsection.
Also, as with the correlation functions, the oscillations have twice the periodicity of the underlying correlation function (again, as nodes occur with double the frequency of a sinusoidal wave).
The mean saddle density for several choices of $a$ is plotted in Fig.~\ref{fig:wall}a.

As shown in Ref.~\onlinecite{ber02c}, when $a \ll 1,$ there is a large additional peak in the vortex density for small $R.$
Robin boundary conditions in this regime are known to have unusual properties, such as admitting negative energy solutions \cite{sie95}, and diverging momenta on the boundary \cite{ber08a}.
When $a$ is small, the peak occurs in the neighborhood $Y \approx a + \varepsilon a^3,$ where $\rho_{\mathrm{v}}(Y;a)$ has the skewed-Lorentzian form \cite{ber02c}
\begin{equation}
     \rho_{\mathrm{v}}(Y = a + a^3 \varepsilon; a) \approx \frac{12\sqrt{1+ 4 (6\varepsilon - 1)^2}}{a^3(1+(12\varepsilon - 1)^2)^{3/2}}
     \label{eq:rhovsmalla},
\end{equation}
with a peak near $Y = a + a^3/12$ of approximately $24\sqrt{2}/ a^3.$
Analysis of Eq.~(\ref{eq:rhosba}) reveals that the saddles, too, have a peak for small $a,$ and an analogous argument as for vortices gives for $a \ll 1,$
\begin{equation}
     \rho_{\mathrm{v}}(Y = a + a^3 \varepsilon; a) \approx \frac{12\sqrt{2+24\varepsilon (6\varepsilon - 1)}}{a^3(1 + 4(6\varepsilon - 1)^2)^{3/2}}
     \label{eq:rhossmalla},
\end{equation}
that is, a peak of almost the same shape and magnitude as for vortices, but with maximum near $Y = a + a^3/6.$
This peak ensures that small $a$ does not give rise to an accumulation of total Poincar\'{e} index near the boundary.
Plots of the $\rho_{\mathrm{v}}(Y;a), \rho_{\mathrm{s}}(Y;a)$ for $a = \pi/10$ are shown in~\ref{fig:wall}c, along with the corresponding approximations.

\section{Comparison between experiment and RWM predictions}
\label{ch::comparison}

In this section we compare the results of our microwave billiard experiment, outlined in section~\ref{ch::exp} with the theoretical predictions of the isotropic and boundary-adapted RWMs described in Section~\ref{ch::Averages}.
Since distances $R, Y$ are scaled with respect to $k$ as in Eqs.~(\ref{eq:Rdef}), (\ref{eq:XYdef}), the experimental results for different frequencies $\nu$ have been superimposed, improving the statistics.

\begin{figure}
\includegraphics[width=8cm]{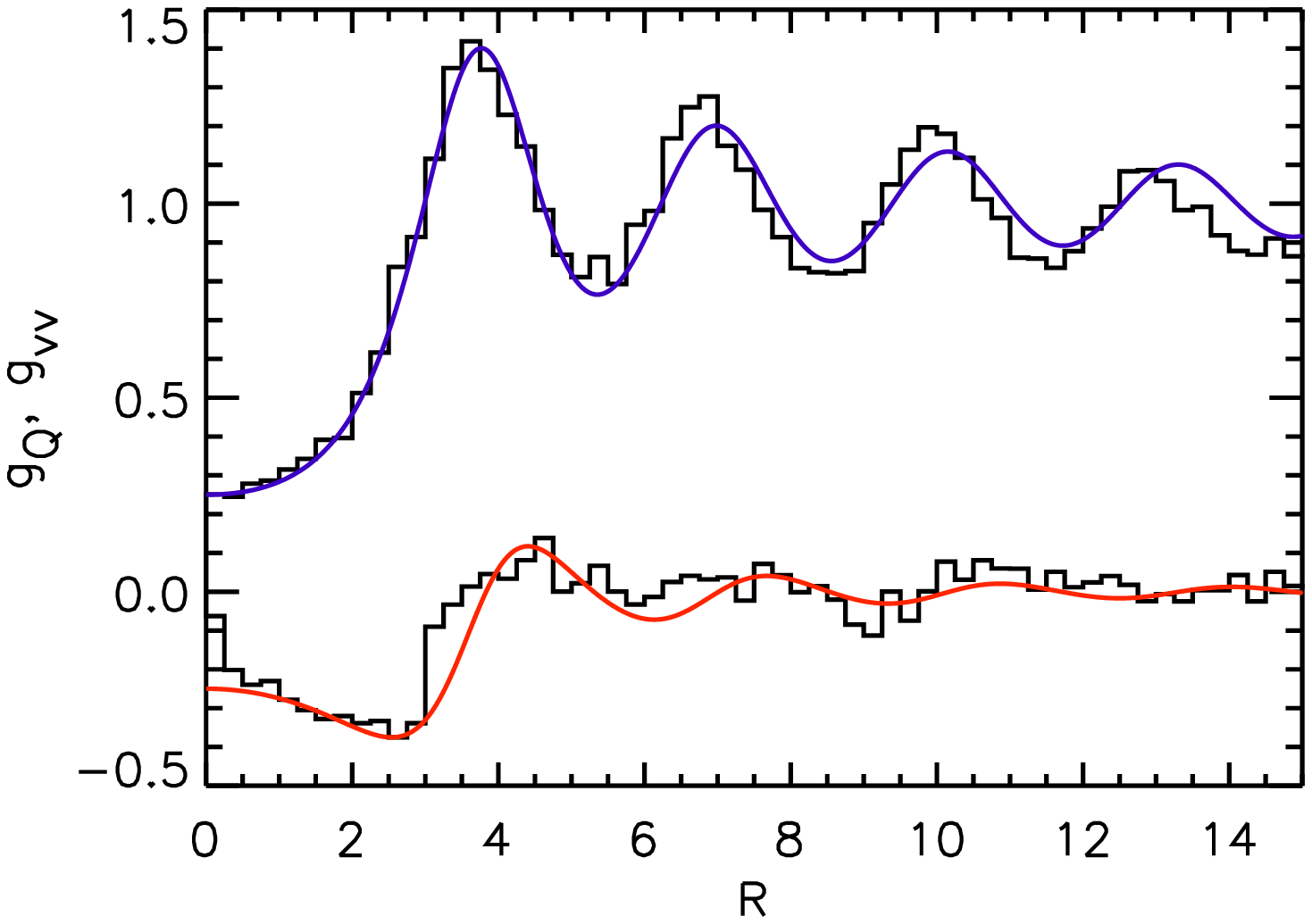}
\includegraphics[width=8cm]{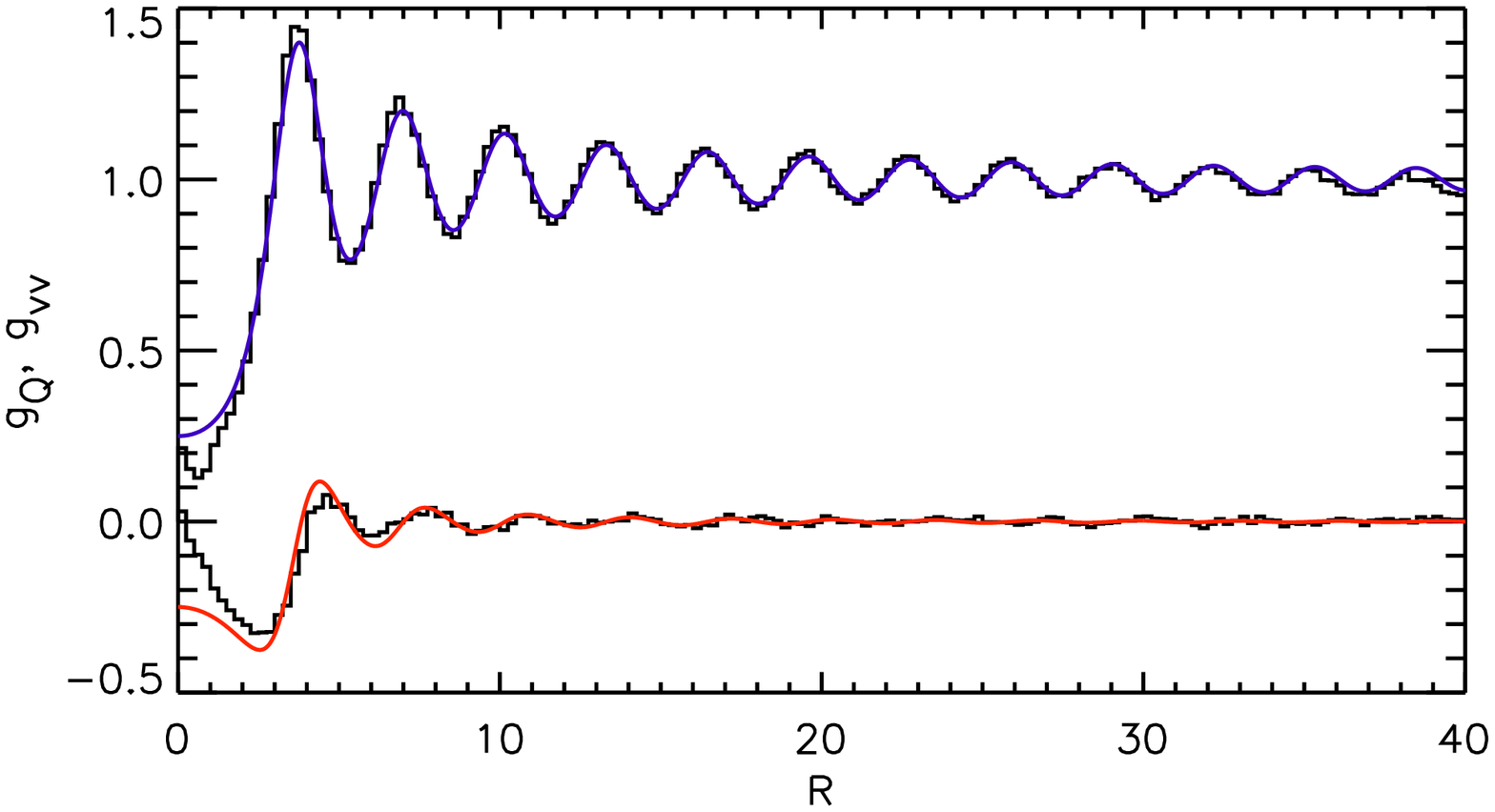}
\caption{\label{fig::paircorrVortex}  (color online)
 Vortex pair correlation $g_{\mathrm{vv}}(R)$ and charge correlation function $g_{Q}(R),$ in (a) the low frequency regime ($\unit[5]{GHz} < \nu < \unit[9]{GHz}$), and (b) the higher frequency regime ($\unit[15]{GHz} < \nu < \unit[18.6]{GHz}$).
 The solid lines correspond to the analytic prediction of RWM discussed in subsection \ref{sub:bulk}, using the exact formula for $g_{\mathrm{vv}}(R)$ rather than the asymptotic form.
}
\end{figure}

Fig.~\ref{fig::paircorrVortex} shows the comparison between experiment and theory for the vortex-vortex pair correlation functions, both signed (Eq.~\ref{eq:gQ}) and unsigned (Ref.~\onlinecite{den01c}, Eqs.~(32)-(35)), in two different frequency regimes.
Because of the limited resolution due to the measurement grid in the low frequency regime the correlation function could only be determined reliably for small $kr,$ see Fig.~\ref{fig::paircorrVortex}a.
But in this regime the system size is comparable to the vortex spacing, leading to an influence of the boundary on the measured correlation function and restricting the observable $R$ range up to $R_{\mathrm{max}}=k L \approx 15$, where $L$ corresponds to a typical length of the system.
Due to the boundary effects the oscillation period of the experimental results is shorter than the theoretical one for the bulk statistics \cite{baec02d,eck99}.
The results for the higher frequency regime, on the other hand, shown in Fig.~\ref{fig::paircorrVortex}b agrees perfectly with theory for large $R$, but fails for small $R$ because of the mentioned limited resolution in the measurement.
Experimental results for the vortex pair correlation function have been published already previously, though with a by far poorer statistics \cite{bar02}.
All other quantities shown in this section have not been published previously.

\begin{figure}
 \includegraphics[width=7.5cm]{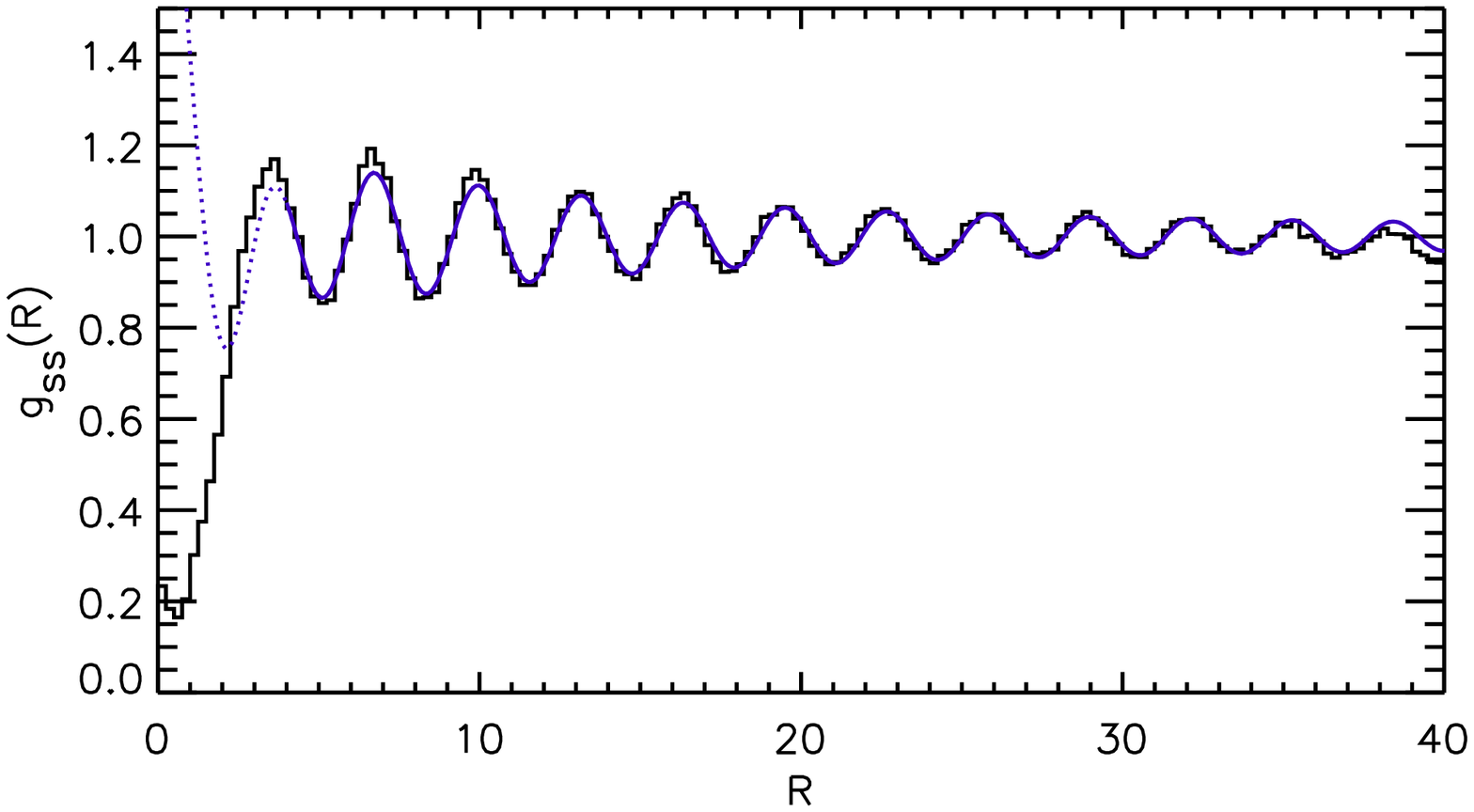}
 \includegraphics[width=7.5cm]{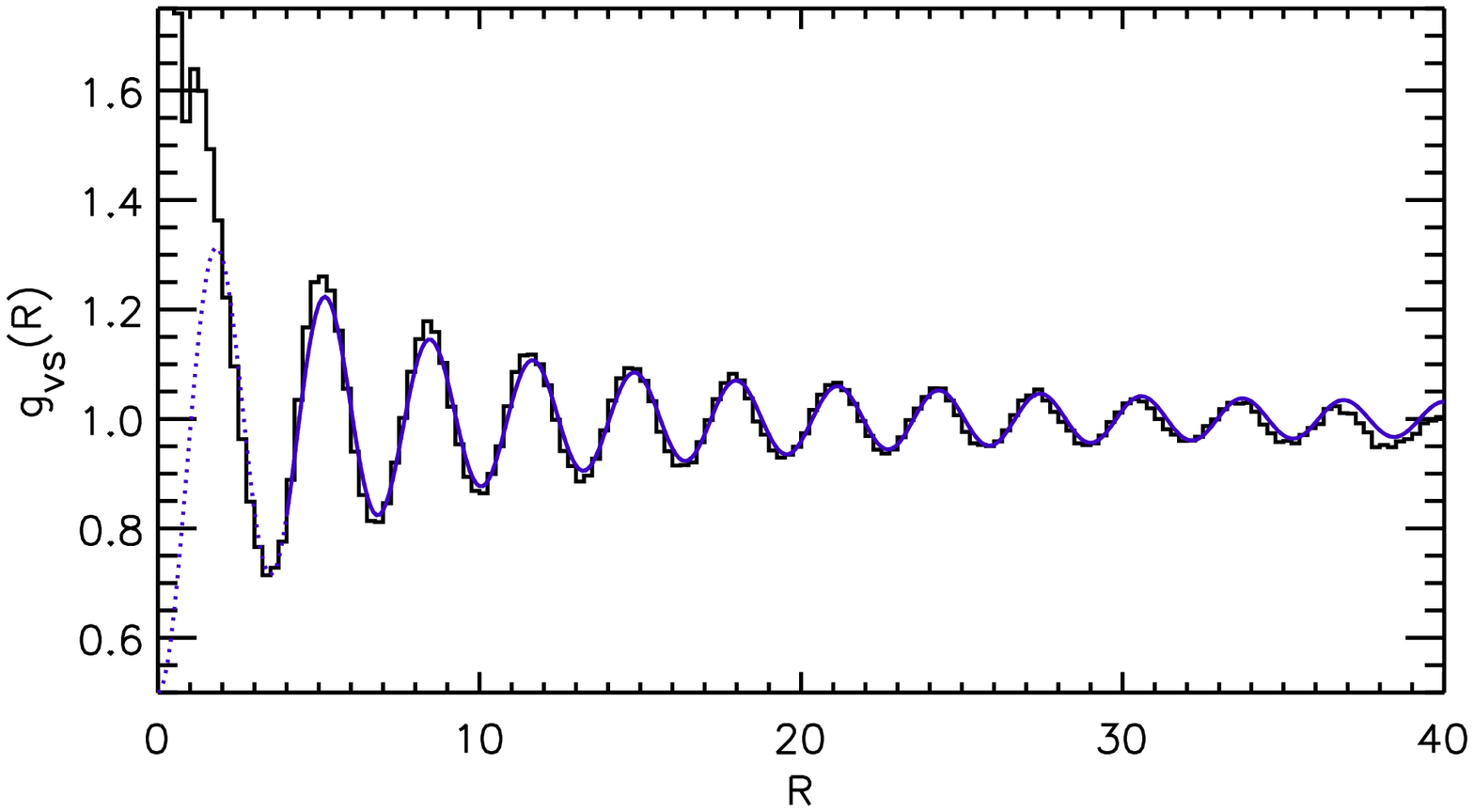}
 \caption{\label{fig::paircorrSaddle} (color online)
 Pair correlations involving saddle points: (a) saddle-saddle correlation function $g_{\mathrm{ss}}(R);$ (b) vortex-saddle function $g_{\mathrm{vs}}(R).$
 Experimental data is plotted against the asymptotic forms of Eqs.~(\ref{eq:asymgvs}), (\ref{eq:asymgss}).}
\end{figure}

Results for the saddle pair correlation function $g_{\mathrm{ss}}(R)$ and the vortex-saddle function $g_{\mathrm{vs}}(R)$ are plotted in Fig.~\ref{fig::paircorrSaddle}a, with experimental data plotted against the asymptotic formulas (\ref{eq:asymgss}), (\ref{eq:asymgvs}).
As with the vortex correlation functions, the agreement between theory and experiment is very good for a wide range of $R$; since the theoretical formulas are asymptotic, we do not have a theory to compare with the data for small $R$.
Fig.~\ref{fig::paircorrSaddle}b shows the total and signed critical point correlation functions $g_{\mathrm{crit}}(R)$
and $g_{\mathrm{I}}(R),$ again with good agreement.

\begin{figure}
 \includegraphics[width=7.5cm]{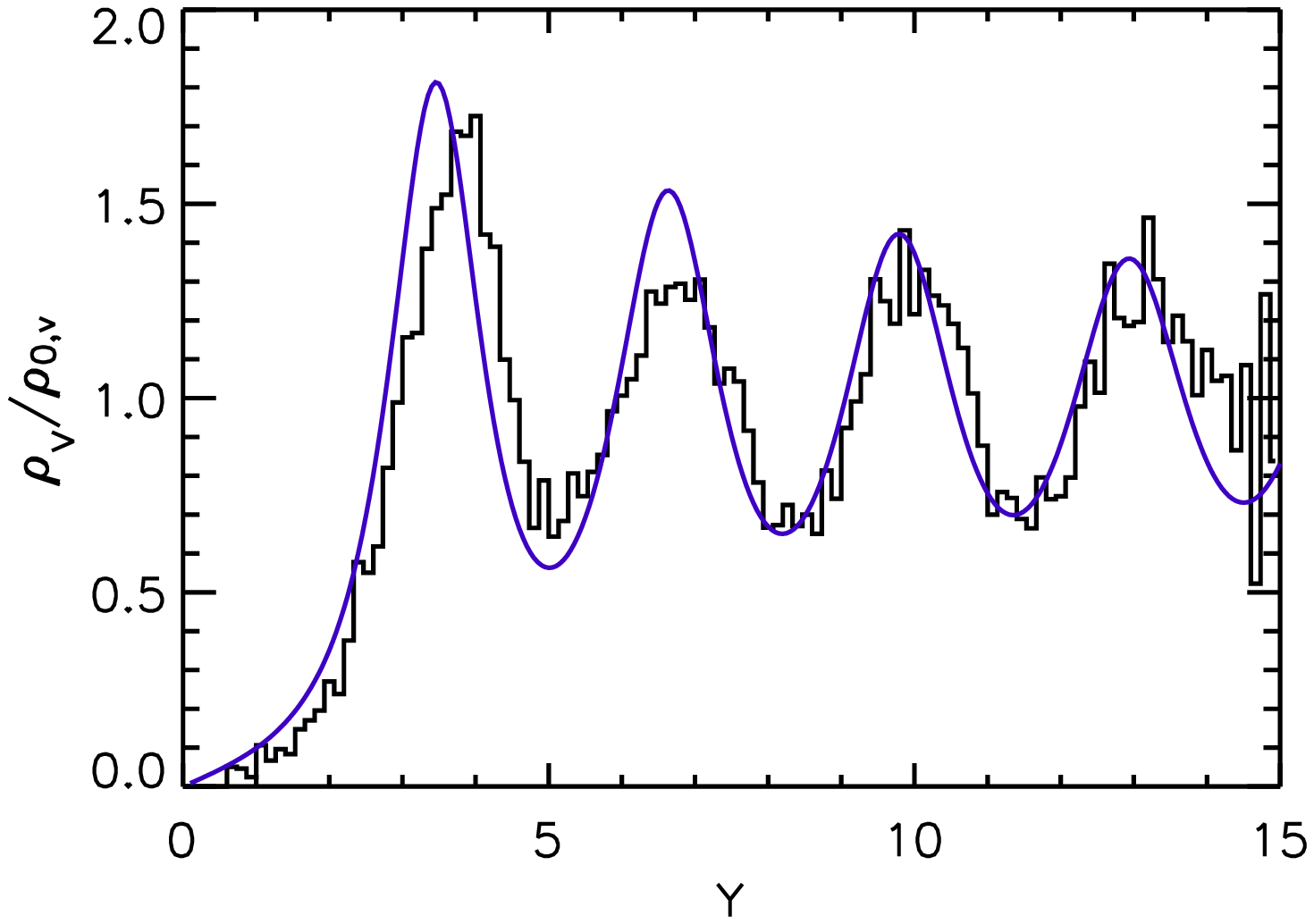}
 \includegraphics[width=7.5cm]{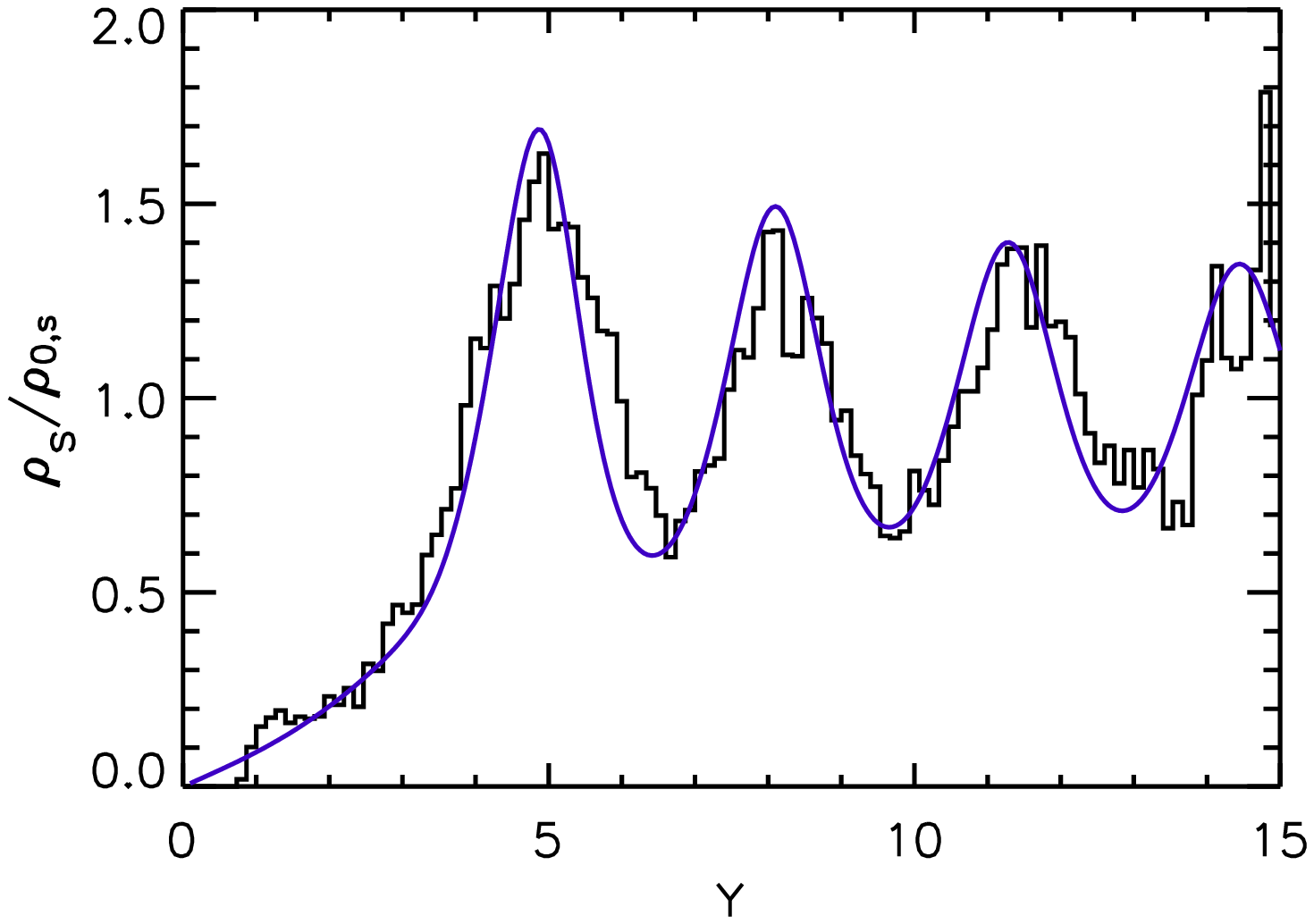}
 \caption{\label{fig:wallplot} (color online)
 Critical point density fluctuations as a function of scaled distance $Y$ from a straight wall satisfying Dirichlet boundary conditions: (a) vortex density; (b) saddle density.
 Experimental data is plotted against the analytic forms of Eqs.~(\ref{eq:rhovba}), (\ref{eq:rhosba}), for Dirichlet boundary conditions.}
\end{figure}

Experimental measurements of the average vortex and saddle density fluctuations against a straight boundary satisfying Dirichlet conditions are shown in Fig.~\ref{fig:wallplot}, vortex fluctuations in the upper panel (with theoretical density from Eq.~(\ref{eq:rhovba})), the lower panel the saddle fluctuations (with theoretical plot from Eq.~(\ref{eq:rhosba})).
The agreement between theory and experiment is excellent.

In summary, by applying a higher grid resolution and a bilinear interpolation technique the flow pattern through an open microwave billiard could
be resolved  by nearly one order of magnitude better as in previous experiments.
This allowed the determination of various distribution and correlation functions for the critical points in the flows, vortices and saddles, which had been inaccessible hitherto.

From the theoretical side, we have presented the universal predictions of the Random Wave Model for the pair correlations of the current's critical points in the bulk, showing excellent agreement with the measurements.
Although some of the results can be obtained in closed form, an asymptotic method valid for large separations is necessary to construct some important correlations.
By power counting of the characteristic decay of correlations with respect to the separation screening can be explicitly checked, leading to a surprising behavior of the Poincar\'e index, which is still to be explained but it is fully supported by the experimental results.

We also contrast for the first time the basic assumptions of the boundary-adapted Random Wave Model against experimental results.
The influence of the boundary showed up first in an oscillatory behavior in the density of vortices and saddles close to the wall.

\appendix

\section{Asymptotic pair correlations}
\label{app:asymptotic}

In this section we sketch the method we use to calculate the asymptotic (large $R$) approximation to the RWM average of an arbitrary functional $\mathcal{F}$ depending on the field and its derivatives at points $\vec{r}_{1}, \vec{r}_{2}$ with $k|\vec{r}_{1}- \vec{r}_{2}|=R$ the scaled distance.

We start with the exact expression for the Gaussian average
\begin{equation}\label{eq:gaussa}
  \langle \mathcal{F} \rangle
  = \frac{1}{\sqrt{(2 \pi)^{n+m} \det{\bf M}}}\int_{-\infty}^{\infty}
  \mathcal{F}[\vec{u}] {\rm e}^{-\frac{1}{2} \vec{u}\cdot{\bf M}^{-1}\cdot \vec{u}} \mathrm{d}^{n+m}\vec{u}.
\end{equation}
where the vector
\begin{equation}
  \vec{u} = (\vec{u}^{(1)},\vec{u}^{(2)}),
\end{equation}
comprises all the relevant degrees of freedom $\vec{u}^{(1)}=(u^{(1)}_{1},\ldots,u^{(1)}_{n})$ at position $\vec{r}_{1}$ and $\vec{u}^{(2)}=(u^{(2)}_{1},\ldots,u^{(2)}_{m})$ at position $\vec{r}_{2}$.
With this definition, the correlation matrix has a natural block form
\begin{equation}
  {\bf M} =\left( \begin{array}{cc}  {\bf M}^{(1,1)} & {\bf M}^{(1,2)} \\  {\bf M}^{(2,1)} & {\bf M}^{(2,2)} \end{array} \right)
  \label{eq:Mblockform}
\end{equation}
where $\left[{\bf M}^{(\alpha,\beta)}\right]_{i,j}=\langle u^{(\alpha)}_{i} u^{(\beta)}_{j} \rangle$.
The key step of the method is to observe that the only dependence of the average $\langle \mathcal{F} \rangle$ on the scaled distance $R$ comes from the off-diagonal blocks, and the known asymptotic expansion of ${\bf M}^{(1,2)}(R)$ will imply an asymptotic expansion of $\langle \mathcal{F} \rangle$.
In order to follow this program, we found convenient to switch to the Fourier representation of the probability distribution in Eq.~(\ref{eq:gaussa})
\begin{eqnarray}
\label{eq:probab}
  \frac{{\rm e}^{-\frac{1}{2} \vec{u}\cdot{\bf M}^{-1}\cdot \vec{u}}}{\sqrt{(2 \pi)^{n+m} \det
{\bf M}}}
  & = &\int_{-\infty}^{\infty}{\rm d}^{n}\vec{w}^{(1)}{\rm d}^{m}\vec{w}^{(2)}{\rm e}^{i(\vec{w}^{(1)},\vec{w}^{(2)})\cdot(\vec{u}^{(1)},\vec{u}^{(2)})}
  \nonumber \\
  & \times & {\rm e}^{-\frac{1}{2}(\vec{w}^{(1)},\vec{w}^{(2)})\cdot{\bf M}\cdot(\vec{w}^{(1)},\vec{w}^{(2)})}.
\end{eqnarray}
Due to the asymptotic form of the off-diagonal block ${\bf M}^{(1,2)} \sim 1/\sqrt{R},$ the last line in Eq.~(\ref{eq:probab}) can be written as
\begin{eqnarray}
  \label{eq:asympa}
  &&{\rm e}^{-\frac{1}{2}(\vec{w}^{(1)},\vec{w}^{(2)})\cdot{\bf M}\cdot(\vec{w}^{(1)},\vec{w}^{(2)})}= \\
  &&{\rm e}^{-\frac{1}{2}\vec{w}^{(1)}\cdot{\bf M}^{(1,1)}\cdot\vec{w}^{(1)}}{\rm e}^{-\frac{1}{2}\vec{w}^{(2)}\cdot{\bf M}^{(2,2)}\cdot\vec{w}^{(2)}} \times \nonumber \\
  && \left[1-\vec{w}^{(1)}\cdot{\bf M}^{(1,2)}\cdot\vec{w}^{(2)}+\frac{1}{2}\left(\vec{w}^{(1)}\cdot{\bf M}^{(1,2)}\cdot\vec{w}^{(2)}\right)^{2}\right]\nonumber \\
  && + O\left(1/R^{3/2}\right). \nonumber
\end{eqnarray}
We note that, given the particular form of the asymptotic expansion of the Bessel function, the very first term of the asymptotic expansion of the off-diagonal blocks not only gives the leading order term in $1/\sqrt{R}$ of the average, but also the subleading one of order $1/R$.
Beyond the subleading order, higher order terms of the average require higher order terms in the expansion of ${\bf M}^{(1,2)}$.
This has to be consider a very special property of the Bessel-correlated RWM with its characteristic slow ($\sim 1/\sqrt{R}$) decay of correlations.

Substitution of Eq.~(\ref{eq:asympa}) into Eq.~(\ref{eq:probab}) results in the asymptotic expansion of the probability distribution which in turn leads to the sought asymptotic expansion of the average in Eq.~(\ref{eq:gaussa}).
The calculations are simple but tedious, so we merely quote the result.
Denoting
\begin{equation}
  {\bf A} = \left[{\bf M}^{(1)}\right]^{-1}{\bf M}^{(1,2)}\left[{\bf M}^{(2)}\right]^{-1}
\end{equation}
and introducing the tensors
\begin{eqnarray}
  F_{0}&=&\langle \mathcal{F} \rangle_{0} \\
  {\bf F}^{(\alpha,\beta)}_{0}&=&\langle u^{(\alpha)}_{i}u^{(\beta)}_{j} \mathcal{F} \rangle_{0} \nonumber
\end{eqnarray}
where $\langle \ldots \rangle_{0}$ indicates the average in Eq.~(\ref{eq:gaussa}) with ${\bf M}^{(1,2)}={\bf M}^{(2,1)}=0$, we get finally
\begin{eqnarray}
  \label{eq:fina}
  \langle \mathcal{F} \rangle &=&F_{0}+{\rm Tr }{\bf A} {\bf F}^{(2,1)}_{0} \nonumber \\
  &+&\frac{1}{2}F_{0}{\rm Tr }{\bf A} {\bf M}^{(2,2)} {\bf A}^{\tau} {\bf M}^{(1,1)} \\
  &-&\frac{1}{2}\left[{\rm Tr }{\bf F}^{(1,1)}_{0}{\bf A} {\bf M}^{(2,2)} {\bf A}^{\tau}
  +{\rm Tr }{\bf F}^{(2,2)}_{0}{\bf A}^{\tau} {\bf M}^{(1,1)} {\bf A}\right] \nonumber \\
  &+&\frac{1}{2}{\rm Tr }{\bf A}{\bf F}^{(2,2)}_{0}{\bf A}^{\tau}{\bf F}^{(1,1)}_{0} +O\left(1/R^{3/2}\right). \nonumber
\end{eqnarray}
Our result Eq.~(\ref{eq:fina}) allow us to calculate the leading ($\sim 1/\sqrt{R}$) and subleading ($\sim 1/R$) contributions to any correlation in terms of the ${\bf M}^{(1,2)}={\bf M}^{(2,1)}=0$ (uncorrelated) results.
It also provides the large $R$ asymptotics to one-point functionals at points separated a distance $2R$ from an infinite straight boundary.

\section{Derivation of the $Y$-dependent saddle density Eq.~(\ref{eq:rhosba})}
\label{app:dervrhosba}

The saddle density at scaled distance $Y$ from a straight boundary can be calculated by standard methods of Gaussian integration (as used, for instance, in Refs.~\onlinecite{den03,den07}), although the details are rather tedious and only outlined here.
Normalized by the bulk density, the $Y$-dependent saddle density is
\begin{eqnarray}
  \rho_{\mathrm{s}}(Y) = 4\pi \langle D_{\mathrm{Y}}(Y) \rangle
   =  \frac{1}{\pi} \int {\rm d}^2\vec{t} \langle \exp( {\rm i} \vec{j}\cdot \vec{t}) \mathcal{J}_{\mathrm{s}} \rangle.
  \label{eq:rhos}
\end{eqnarray}
Since $\mathcal{J}_{\mathrm{s}}$ is a sum of terms, linearity of the average simplifies the Gaussian average to calculations of the form
\begin{equation}
  A_{\alpha\beta} = \langle (\xi \eta_{\alpha\beta} - \eta \xi_{\alpha\beta})^2 \exp( {\rm i} \vec{j}\cdot \vec{t}) \rangle,
  \label{eq:Aab}
\end{equation}
where $\alpha\beta = XX, YY$ or $XY.$

Each $A_{\alpha\beta}$ is an average over an 8-dimensional Gaussian random vector
\begin{equation}
  \vec{u} = \{\xi,\xi_X,\xi_Y,\xi_{\alpha\beta},\eta,\eta_X,\eta_Y,\eta_{\alpha\beta}\}.
  \label{eq:udef}
\end{equation}
The terms in $\xi$ and $\eta$ are uncorrelated.
Denoting either by $v,$ and suppressing $Y$-dependence, the relevant non-vanishing correlations follow from Eq.~(\ref{eq:Bdef}), $B = \langle v^2 \rangle$ :
\begin{eqnarray}
  \langle v v_{YY} \rangle & = & \tfrac{1}{2} - \tfrac{1}{4} B'', \nonumber \\
  \langle v_X^2 \rangle & = & - \tfrac{1}{2} + B + \tfrac{1}{4} B'' , \nonumber \\
  \langle v_Y^2 \rangle & = & \tfrac{1}{2} + \tfrac{1}{4} B'', \nonumber \\
  \langle v v_Y \rangle & = & \tfrac{1}{2} B', \nonumber \\
  \langle v_Y v_{XX} \rangle & = & -\tfrac{1}{2} B' - \tfrac{1}{8}B''', \nonumber \\
  \langle v_Y v_{YY} \rangle & = & \tfrac{1}{8}B''',\nonumber \\
  \langle v_{XX}^2 \rangle & = & -\tfrac{5}{8} + B + \tfrac{1}{2} B'' + \tfrac{1}{16} B'''',  \nonumber \\
  \langle v_{YY}^2 \rangle & = & \tfrac{3}{8} + \tfrac{1}{16} B'''',\nonumber \\
  \langle v_{XY}^2 \rangle & = & \tfrac{1}{8} + \tfrac{1}{4} B'' + \tfrac{1}{16} B'''',
  \label{eq:corrs1}
\end{eqnarray}
where the prime denotes the partial derivative with respect to $Y$.
We denote the appropriate correlation matrix for $\vec{u}$ by $\mathbf{M}_{\vec{u}}.$

The scalar product in the exponent in (\ref{eq:Aab}) can be written as a symmetric quadratic form in $\vec{u},$
\begin{equation}
  \vec{j}\cdot \vec{t} = \tfrac{1}{2} \vec{u}\cdot \mathbf{T} \cdot \vec{u},
  \label{eq:Tdef}
\end{equation}
where $\mathbf{T}$ depends on $t_1, t_2.$

Therefore, defining the matrix $\boldsymbol{\Xi} \equiv (\mathbf{M}_{\vec{u}}^{-1} +  {\rm i} \mathbf{T})^{-1},$ and $Q \equiv (\xi \eta_{\alpha\beta} - \eta \xi_{\alpha\beta}),$ it is straightforward to see
\begin{eqnarray}
  A_{\alpha\beta} & = & \frac{1}{(2\pi)^4\sqrt{\det\mathbf{M}_{\vec{u}}}} \int {\rm d}^8 \vec{u} Q^2 \exp(-\tfrac{1}{2} \vec{u}\cdot \boldsymbol{\Xi}^{-1}\cdot\vec{u}) \nonumber \\
  & = & \sqrt{\frac{\det\boldsymbol{\Xi}}{\det\mathbf{M}_{\vec{u}}}} \left[ \Delta^2 \exp(-\tfrac{1}{2}\vec{w}\cdot\boldsymbol{\Xi}\cdot\vec{w})\right]_{\vec{w}  =0}
  \label{eq:Aab2}
\end{eqnarray}
where $\vec{w}$ is a Fourier dual to $\vec{u},$ and $\Delta$ is the quadratic form of Fourier derivatives corresponding to $Q.$
The final step in Gaussian integration by parts reduces to a multilinear combination of entries of $\boldsymbol{\Xi}.$
Each $A_{\alpha\beta}$ can now be integrated with respect to $\vec{t}.$
In terms of the original correlations, the final result is
\begin{eqnarray}
  \rho_{\mathrm{s}}(Y) & = & \frac{\sqrt{B}}{\langle v_X^2 \rangle^{3/2}(B \langle v_Y^2 \rangle - \langle v v_Y \rangle^2)^{3/2} } \left[ (B \langle v_Y^2 \rangle - \langle v v_Y \rangle^2)\right.\nonumber \\
  & &  (\langle v_X^2 \rangle (\langle v_{XX}^2 \rangle + \langle v_{YY}^2 \rangle + 2 \langle v_{XY}^2 \rangle) - 2 \langle v_Y v_{XX} \rangle^2) \nonumber \\
  & & - \langle v_X^2 \rangle \langle v_Y^2 \rangle (\langle v_X^2 \rangle^2 + \langle v v_{YY} \rangle^2)  \nonumber \\
  & & - B \langle v_X^2 \rangle (\langle v_Y v_{XX} \rangle^2 + \langle v_Y v_{YY} \rangle^2)
  \label{eq:rhosfinal} \\
  & & \left. - 2 \langle v_X^2 \rangle \langle v v_Y \rangle (\langle v_X^2 \rangle \langle v_Y v_{XX} \rangle + \langle v v_{YY} \rangle \langle v_Y v_{YY} \rangle)\right]. \nonumber
\end{eqnarray}
Eq.~(\ref{eq:rhosba}) follows from this expression with the appropriate substitutions from Eqs.~(\ref{eq:corrs1}).

%

\acknowledgments This work was founded by the Deutsche Forschungsgemeinschaft via the Forschergruppe 760 ``Scattering Systems with Complex Dynamics''.
MRD is supported by the Royal Society of London.


\end{document}